\documentclass[oldversion]{aa}
\usepackage{natbib}
\bibpunct{(}{)}{;}{a}{}{,}
\usepackage{graphicx}
\usepackage{epsf}
\usepackage{rotating}
\usepackage{graphics}
\usepackage{txfonts}

\newcommand{\MJ}{M$_{Jup}$}
\newcommand{\RJ}{R$_{Jup}$}
\newcommand{\RS}{R$_{\odot}$}
\newcommand{\kms}{km\,s$^{-1}$\/}
\newcommand{\ms}{m\,s$^{-1}$\/}
\begin{document}
   \title{Refined parameters and spectroscopic transit of the super-massive planet HD147506b
    \thanks{Based on observations collected with the SOPHIE spectrograph on the 1.93-m telescope at
    OHP, France (programs 07A.PNP.MAZE and 07A.PNP.CONS)}}

   \author{B. Loeillet\inst{1,2}
      \fnmsep
      \and
      A.~Shporer\inst{3}
      \and
      F.~Bouchy\inst{2}
      \and
      F.~Pont\inst{4}
      \and
      T.~Mazeh\inst{3}
      \and
      J.L.~Beuzit\inst{5}
      \and
      I.~Boisse\inst{2}
      \and
      X.~Bonfils\inst{5}
      \and
      R.~Da Silva\inst{4}
      \and
      X.~Delfosse\inst{5}
      \and
      M.~Desort\inst{5}
      \and
      A.~Ecuvillon\inst{2}
      \and
      T.~Forveille\inst{5}
      \and
      F.~Galland\inst{5}
      \and
      A.~Gallenne\inst{2}
      \and
      G.~H\'ebrard\inst{2}
      \and
      A.-M. Lagrange\inst{5}
      \and
      C.~Lovis\inst{4}
      \and
      M.~Mayor\inst{4}
      \and
      C.~Moutou\inst{1}
      \and
      F.~Pepe\inst{4}
      \and
      C.~Perrier\inst{5}
      \and
      D.~Queloz\inst{4}
      \and
      D.~S\'egransan\inst{4}
      \and
      J.P. Sivan\inst{1}
      \and
      N. C. Santos\inst{4,6}
      \and
      Y.~Tsodikovich\inst{3}
      \and
      S.~Udry\inst{4}
      \and
      A.~Vidal-Madjar\inst{2}
      }

   \offprints{B. Loeillet,
     \email benoit.loeillet@oamp.fr}

   \institute{Laboratoire d'Astrophysique de Marseille, BP 8, 13376 Marseille
     cedex 12, France, Universit\'e de Provence, CNRS (UMR 6110) and CNES
          \email{benoit.loeillet@oamp.fr}
     \and
         Institut d'Astrophysique de Paris, UMR7095 CNRS, Universite Pierre et Marie Curie, 98$^{bis}$ Bd Arago, 75014 Paris, France
     \and
         Wise Observatory, Tel Aviv University, Israel 69978
     \and
         Observatoire de Gen\`eve, Universit\'e de Gen\`eve, 51 ch. des Maillettes, 1290 Sauverny, Switzerland
     \and
         Laboratoire d'Astrophysique de Grenoble, Observatoire de Grenoble, UMR5571 Universit\'e
	 J. Fourier et CNRS, BP 53, 38041 Grenoble, France
     \and
         Centro de Astrofísica, Universidade do Porto, Rua das Estrelas, 4150-762 Porto, Portugal
             }

   \date{}

%abstract

  \abstract
  {In this paper, we report a refined determination of the orbital parameters
  and the detection of the Rossiter-McLaughlin effect of the recently discovered
  transiting exoplanet HD147506b (HAT-P-2b). The large orbital eccentricity at the short orbital
  period of this exoplanet is unexpected and is distinguishing from other known 
  transiting exoplanets.
  We performed high-precision radial velocity spectroscopic observations 
  of HD147506 (HAT-P-2) with the new spectrograph SOPHIE, mounted on the 1.93~m telescope at the
  Haute-Provence observatory (OHP).
  We obtained 63 new measurements, including 35 on May 14 and 20 on June 11, 
  when the planet was transiting its parent star. 
  The radial velocity (RV) anomaly observed illustrates that HAT-P-2b orbital motion is set in the 
  same direction as its parent star spin.
  The sky-projected angle between the normal of the orbital plane and the stellar spin 
  axis, ${\bf \lambda = 0.2^{+12.2}_{-12.5}}$~\degr, is consistent with zero. The planetary and stellar 
  radii were re-determined, yielding ${\bf R_p = 0.951^{+0.039}_{-0.053}}$ \RJ\/, 
  ${\bf R_s = 1.416^{+0.040}_{-0.062}}$ \RS.
  The mass (${\bf M_p = 8.62^{+0.39}_{-0.55}}$ \MJ) and radius of HAT-P-2b
  indicate a density of ${\bf 12.5^{+2.6}_{-3.6}}$~g cm$^{-3}$, 
  suggesting an object in between the known close-in planets with typical density of 
  the order of 1 g cm$^{-3}$, and the very low-mass stars, with density greater than 
  50~g~cm$^{-3}$.
  }


  \authorrunning{Loeillet B. et al.}
  \titlerunning{Refined parameters and spectroscopic transit of HD147506b}
  \maketitle

%
%________________________________________________

%=======================================================================================
\section{Introduction}
%=======================================================================================
Thirty of the almost 270 known extrasolar planets have been observed to transit
their parent stars\footnote{see
http://obswww.unige.ch/$\sim$pont/TRANSITS.htm}. This small subgroup of
planets have the highest impact on our overall understanding of
close-in giant planets because we can estimate their mass and radius,
and for some of them directly study their atmosphere. These
transiting hot Jupiters have masses from 0.07 to about 8 {\MJ} and
radii from 0.3 to about 1.4 \RJ. This set of planets was recently
extended to hot Neptune objects by the discovery of the transit of
GJ436b \citep{Gillon07}, and to super-massive planets by the
detection of HD147506b (HAT-P-2b) by \cite{Bakos07} (hereafter B07). The
discovery paper of the latter derived the key orbital and physical
parameters of this exceptional object, which differs by its mass (9.04 \MJ), orbital 
period (5.63 days), and eccentricity (0.52) from other transiting planets. The radius of HD147506b implies 
an uncommon measured mean density (11.9 g\,cm$^{-3}$) and surface gravity (227 m\,s$^{-2}$). These properties
suggest that HAT-P-2b might be an intermediate object between giant planets and low-mass stars, 
near the brown dwarf population. Its density is indeed close to the upper limit of the planetary models
\citep{Baraffe03} and may put this object in the transition region between planets and brown dwarfs.

We report here new RV measurements of HD147506 (HAT-P-2)
obtained to provide additional information and constraints on
this unusual planetary system. This was done by 1) refining the
orbital parameters, 2) refining the mass and radius of the companion
and 3) measuring and modeling the Rossiter-McLaughlin effect (RV anomaly
due to the partial eclipse of the rotating stellar surface). We present here a more precise orbital solution and the measurement of the sky projection of the inclination between the stellar spin axis and the normal of the orbital plane.

%=======================================================================================
\section{Observations}
%=======================================================================================
The parent star of HAT-P-2b, HD147506, was observed in May and June 2007 with
the new spectrograph SOPHIE \citep{Bouchy06} mounted on the
1.93-m telescope of Haute Provence Observatory. Observations were
conducted in the high-efficiency mode (HE mode), which provides a spectral resolution of  R$\sim$~39,000. 
The Thorium-Argon lamp was used to calibrate the wavelength scale. The
simultaneous ThAr calibration mode was not used, as wavelength calibration was performed less than 1
hour before and after the observations leading to an instrumental stability of less than 2 {\ms}.

We obtained eight out-of-transit spectra on six nights between May 6 and 15, 2007, and
a sequence of 6 hours, comprised of 35 spectra during
and after the transit on May 14 (JD=2454235). Unfortunately, 
we could not observe the ingress phase of the transit that occurred just before
twilight. Another sequence was obtained
during transit on June 11 (JD=2454263) to get a full coverage of the transit.
The typical exposure time was 10 minutes, long enough to
reach a sufficiently high signal-to-noise ratio (SNR) and short enough to adequately sample the
observed transit. The average SNR per pixel, at $\lambda=$5500 $\AA$, was about 70 during the first
observed transit and about 90 during the second one. Another limitation of high-precision RV
measurement in the mode used (HE) concerns the guidance centring of the target in the fiber.
As the diameter of the fiber is quite large (3 arcsec), a strong decentring under very good 
seeing conditions ($<$1.5 arcsec) could indeed induce a RV shift of a few tens {\ms}.

We determined RVs using a weighted cross-correlation method, following the
procedure of \cite{Baranne96} and \cite{Pepe05}, with a numerical
mask constructed from the solar spectrum atlas corresponding to a G2
dwarf star. This standard mask is well adapted to the F8-type spectrum of the primary star.
We estimated the measurement uncertainties based on the
photon-noise empirical relation detailed by \cite{Bouchy05} and
\cite{Cameron07}. For the spectra obtained during the night of the first and the second
transit, the typical photon-noise uncertainty is 14 and 10 {\ms}, respectively, whereas for the other 
8 spectra this uncertainty ranges from 6 to 28 {\ms}. The journal of SOPHIE observations, including barycentric
Julian dates (BJD), RVs, photon-noise uncertainties, and 
SNR per pixel is given in Table~\ref{RVtable} (online material). 
The phase-folded RVs are plotted in Fig.~\ref{orbite}. As illustrated by the higher SNR 
values for the data acquired during the second night of observations, we have strong evidence 
that the observational conditions were very good. As explained above, the data are thus affected 
by guiding noise and we assumed an additional systematic error of 50 {\ms\/}. 
We estimated this systematic error thanks to later instrumental tests under the same conditions.

\begin{table}
\caption{\label {RVtable} SOPHIE Radial velocities of HD147506 (HAT-P-2)}
\begin{center}{
\begin{tabular}{c c c c}\hline
\hline
BJD & RV & Photon-noise & Signal-to-noise ratio\\
-2400000 & [\kms] & uncertainties & per pixel \\
 &  & [\kms] & at $\lambda$=5500\AA\\\hline
54227.5016 & -19.4014 & 0.0088 & 109\\
54227.6000 & -19.4082 & 0.0065 & 146\\
54228.5842 & -19.5581 & 0.0188 & 54\\
54229.5993 & -20.1874 & 0.0161 & 61\\
54230.4475 & -21.2249 & 0.0141 & 68\\
54230.6029 & -20.8536 & 0.0148 & 66\\
54231.5987 & -19.5311 & 0.0121 & 78\\
54235.3466 & -20.1916 & 0.0156 & 61\\
54235.3538 & -20.2318 & 0.0180 & 53\\
54235.3615 & -20.3008 & 0.0167 & 57\\
54235.3692 & -20.2790 & 0.0173 & 55\\
54235.3765 & -20.3083 & 0.0172 & 56\\
54235.3866 & -20.3889 & 0.0209 & 46\\
54235.3938 & -20.4280 & 0.0171 & 56\\
54235.4011 & -20.4370 & 0.0173 & 55\\
54235.4088 & -20.4450 & 0.0163 & 58\\
54235.4161 & -20.4530 & 0.0180 & 53\\
54235.4234 & -20.4987 & 0.0150 & 62\\
54235.4310 & -20.5197 & 0.0144 & 65\\
54235.4383 & -20.5032 & 0.0157 & 60\\
54235.4456 & -20.5318 & 0.0178 & 53\\
54235.4535 & -20.5340 & 0.0182 & 52\\
54235.4608 & -20.4990 & 0.0143 & 66\\
54235.4681 & -20.5047 & 0.0124 & 75\\
54235.4759 & -20.4747 & 0.0117 & 81\\
54235.4831 & -20.4974 & 0.0113 & 82\\
54235.4904 & -20.5364 & 0.0111 & 84\\
54235.4981 & -20.5384 & 0.0111 & 85\\
54235.5054 & -20.5484 & 0.0108 & 87\\
54235.5126 & -20.5785 & 0.0135 & 70\\
54235.5204 & -20.6029 & 0.0142 & 66\\
54235.5277 & -20.6066 & 0.0153 & 62\\
54235.5350 & -20.5946 & 0.0115 & 81\\
54235.5434 & -20.6155 & 0.0118 & 80\\
54235.5507 & -20.5987 & 0.0113 & 82\\
54235.5580 & -20.6266 & 0.0110 & 85\\
54235.5682 & -20.6683 & 0.0119 & 80\\
54235.5755 & -20.6712 & 0.0123 & 77\\
54235.5827 & -20.6585 & 0.0115 & 82\\
54235.5905 & -20.6819 & 0.0108 & 87\\
54235.5978 & -20.7002 & 0.0128 & 74\\
54235.6051 & -20.7205 & 0.0117 & 81\\
54236.5190 & -20.2207 & 0.0056 & 99\\
54263.4521 & -20.2050 & 0.0157 & 71 \\
54263.4594 & -20.1880 & 0.0100 & 102\\
54263.4666 & -20.2048 & 0.0089 & 113\\
54263.4739 & -20.1887 & 0.0092 & 110\\
54263.4804 & -20.1626 & 0.0114 & 90\\
54263.4860 & -20.1750 & 0.0120 & 86 \\
54263.4915 & -20.1832 & 0.0128 & 81 \\
54263.4971 & -20.1429 & 0.0096 & 104\\
54263.5030 & -20.1565 & 0.0102 & 99 \\
54263.5086 & -20.1615 & 0.0108 & 95 \\
54263.5141 & -20.1610 & 0.0104 & 97 \\
54263.5236 & -20.1810 & 0.0103 & 99 \\
54263.5291 & -20.1955 & 0.0151 & 72 \\
54263.5347 & -20.3046 & 0.0179 & 63 \\
54263.5411 & -20.3391 & 0.0277 & 48 \\
54263.5561 & -20.3457 & 0.0127 & 85 \\
54263.5634 & -20.4009 & 0.0133 & 82 \\
54263.5706 & -20.4169 & 0.0173 & 67 \\
54263.5779 & -20.4377 & 0.0107 & 99 \\
54263.5852 & -20.4763 & 0.0091 & 113\\
\hline
\end{tabular}}
\end{center}
\end{table}

%=======================================================================================
\section{Stellar properties}
%=======================================================================================
As described in B07, the spectroscopic determination of the radius of HD147506 
is very sensitive to the method used, as well as to the log~$g$ 
determination, which may be affected by the uncertainty of the
spectrum continuum and the large projected rotational velocity, 
$v\sin{I_s}$ of the star. The spectroscopic approach described in B07 provides a stellar radius 
of 1.474$^{+0.062}_{-0.167}$ \RS. Our combined SOPHIE spectrum provides an independent spectroscopic 
determination of the stellar parameters ($R_S = 1.416 ^{+0.040}_{-0.062}$ \RS), consistent with B07.

>From the full width half maximum (FWHM) of the averaged cross-correlation functions (CCF) of
SOPHIE spectra, which were calibrated to yield stellar $v\sin{I_s}$
values \citep{Santos02}, we determined the $v\sin{I_s}$ of HD147506 to
be equal to 21.3$\pm$1.3~{\kms}, somewhat slightly larger than the velocity derived
by B07 (19.8$\pm$1.6~{\kms}). The metallicity index
[Fe/H]=0.11$\pm$0.10 we obtained, using the method described in \cite{Santos02} is in full agreement with B07. The
activity index of log $R'_{HK}$=-4.75$\pm$0.02 was derived from the H
and K CaII lines and appears to be close to the value
determined by B07 (log $R'_{HK}$=-4.72$\pm$0.05). As estimated in \cite{Santos00},
our activity index for a F8 dwarf star implies a stellar jitter from a few \ms to about 20 {\ms}, which is
confirmed by the calibration made by \cite{Wright05}. 
Analysis of the line-bisector computed for all out-of-transit spectra
does not show significant variations nor correlation with the RVs.

%=======================================================================================
\section{Determination of the planetary system parameters}
%=======================================================================================
In this section, we first describe the procedure used to fit the observed velocities with the Keplerian 
orbit and the RM RV anomaly, and how we estimated the uncertainties. 
In the second and third subsections, we discuss our results.
%---------------------------------------------------------------------------------------
\subsection{Analysis of the RV data}
%---------------------------------------------------------------------------------------
We used all available high-precision spectroscopic data to model the orbit and the RM effect
simultaneously. This data includes $10$ Lick spectra and $13$ Keck spectra obtained by B07, 
and the $63$ SOPHIE spectra. Our model is comprised
of $15$ parameters: The period, $P$; periastron passage time, $T_0$; orbital eccentricity, $e$;
angle between ascending node and periastron, $\omega$; RV semi-amplitude, $K$; RV zero point, 
$V_0$ (these first six are the classical orbital parameters); planetary to stellar radii ratio,
$R_p / R_s$; orbital semi-major axis to stellar radius ratio, $a/R_s$; angle between sky 
projection of the orbital angular momentum axis and stellar spin axis, $\lambda$; line of sight
stellar rotational velocity, $v\sin{I_s}$; orbital inclination angle, $i$; and the stellar linear 
limb darkening coefficient, $\epsilon$. We have also determined a velocity shift between
Keck and Lick velocity zero points, $\Delta v_{KL}$, and SOPHIE and Keck zero points, 
$\Delta v_{SK}$. We also estimated a velocity shift, $\Delta v_{S2}$, between the SOPHIE
measurements taken on the second transit night (June 11) and the rest of the SOPHIE 
measurements, taken about a month earlier. The period was fixed on the value given by B07 
($P=5.63341$ days), considering its very high accuracy ($11 s$) derived from extensive photometric 
observations. A linear 
limb-darkening coefficient $\epsilon$=0.71 was also used, considering 
the stellar T$_{eff} = 6250$ K \citep{Claret04}. Hence, our model has $13$ free parameters,
where $8$ are non-linear ($T_0, e, \omega, R_p / R_s, a/R_s, \lambda, v\sin{I_s}$ and $i$) 
and $5$ linear ($K, V_0, \Delta v_{KL}, \Delta v_{SK}$ and $\Delta v_{S2}$).

We used a Keplerian model for the orbit, and the analytic approach described by \cite{Ohta05}, 
to model the RM effect. Equations given by \cite{Ohta05} for the RM RV anomaly were modified
to make them dependent on $R_p/R_s$ and $a/R_s$, instead of $R_p, R_s$ and $a$.

We searched the parameter space for the global minimum $\chi^2$ position, using the 
equation below. 
We used the linear least square method, along with the data, to calculate the linear parameters 
at each position in the parameter space of the non-linear parameters. 
We modified our $\chi^2$ function to account for external 
information, namely the line-of-sight stellar rotational velocity, derived here from the spectra, the 
radii ratio, and transit duration, from B07:
$$
\chi^2 = \Sigma_i (\frac{RV_{o,i} - RV_{c,i}}{RV_{err,i}})^2 + (\frac{v\sin{I_s} - 21.3}{1.3})^2$$
$$
+ (\frac{R_p/R_s - 0.0684}{0.0009})^2 + (\frac{T_{dur} - 0.177}{0.002})^2,
$$
where $RV_{o,i}$ and $RV_{c,i}$ are the $i$-th observed and calculated RVs and $RV_{err,i}$ is 
its error. $T_{dur}$ is the transit duration and is related to $a/R_s$, $R_p/R_s$, $i$ and the 
orbital parameters $P$, $e$, and $\omega$.
The uncertainties were computed directly from the linear least squares analysis for the linear
parameters \citep[][Sect 15.4, Eq 15.4.15]{NumericalRecipes}. The procedure is a bit more complex concerning 
the non-linear
parameters. For each specific parameter, we increase and decrease it
using small steps starting from the minimum $\chi^2$ solution value. At each step, the rest of the
parameters are fitted while holding the specific parameter constant. Then, a 4th degree polynomial 
is fitted to the $\chi^2$ values obtained and the 1-$\sigma$ uncertainty is estimated by identifying 
the values of the fit corresponding to the minimum $\chi^2$ value $+1$.

In the fit procedure, we adopted the B07 stellar jitter of $60$ {\ms} for the Lick and Keck
measurements. For the 
SOPHIE measurements, we adopted a jitter of $17$ {\ms}, resulting in a reduced $\chi^2 = 1$ for the 
out-of-transit SOPHIE measurements. This value is in a good agreement with the value 
estimated from our revised activity index and should be
compared to the jitter of $60$ {\ms} mentioned in B07, considering the span covered 
by Lick and Keck observations was 240 nights. Over such a long
span, one cannot exclude variations in the stellar activity level. We also note that the effect 
of this jitter may have a time scale comparable or longer than the time of a single exposure, 
inducing a correlated noise effect.

\begin{figure}
\centering{
\includegraphics[scale=0.4]{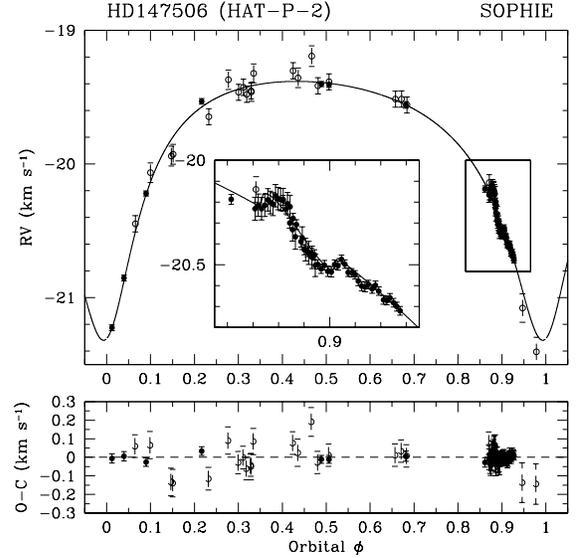}
\caption{\label {orbite} Top: Phase-folded Radial Velocity measurements of
HAT-P-2 superimposed on the refined Keplerian orbital solution.
Open circles refer to the Keck and Lick measurements. Filled circles refer to the SOPHIE
measurements. The inset shows a zoom around the transit where radial velocities exhibit the
Rossiter-McLaughlin effect. 
Bottom: Residuals around the orbital solution.} }
\end{figure}

%---------------------------------------------------------------------------------------
\subsection{Results of the fitted orbital solution}
\begin{figure}
\centering{
\includegraphics[scale=0.4]{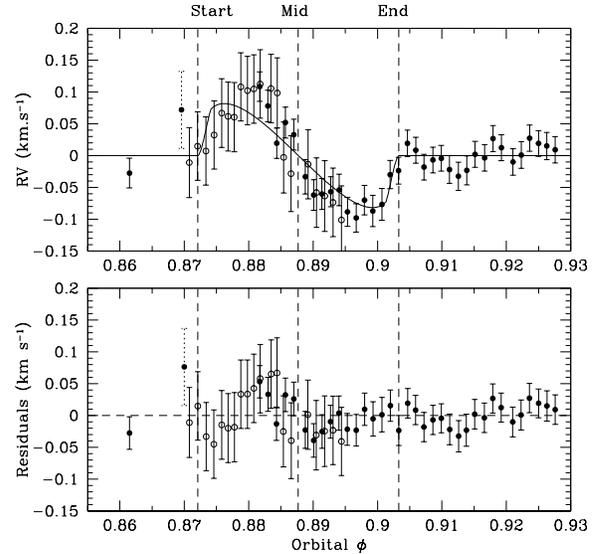}
\caption{\label {rossiterfit} Top: Radial velocities of HAT-P-2 as a function of the orbital phase
after subtraction of the Keplerian model and superimposed with the best fit of the Rossiter-McLaughlin
effect. The filled and open circles represent the RV measurements obtained during the
first and the second sequence of observations of the transit with SOPHIE, respectively. 
The open circle with dotted error bars
represent one measurement from the Keck set which is a few minutes before the ingress phase.
Bottom: RV residuals after subtracting the orbital solution and the modeled RM effect.}
}
\end{figure}

%---------------------------------------------------------------------------------------
Table~\ref{orbittable} lists the result for the fitted parameters of our model. 
The solution depends on the RV data and the prior constraints ($v\sin{I_s}, R_p/R_s$, transit
duration), derived from the 
spectroscopic analysis in this work and the photometry from B07. The refined 
orbital solution is plotted in the upper panel of Fig.~\ref{orbite}, and 
Fig.~\ref{rossiterfit} presents the RM RV anomaly model after subtraction of the 
Keplerian orbit.

The RMS of the out-of-transit residuals of SOPHIE measurements, spanning 36 days, 
is equal to 18~{\ms}, whereas RMS of all out-of-transit measurements, spanning 282 days,
is 50~{\ms}. The orbital parameters we derived are consistent with B07. However our 
uncertainties are smaller, due to a larger sample of high-precision RV data.

We searched for a second planetary signal in the RV 
residuals to look for hints of a third body in the system. No clear periodic signal
appears in the RV residuals. We estimated that we can exclude the presence 
of a second planet of mass greater than 1.3, 1.5 and 1.8\MJ\/ for an orbital period shorter 
than 50, 100, and 200 days respectively. However, the increased RMS for all out-of-transit residuals
suggests that a long-term RV follow-up of this star is needed.

%---------------------------------------------------------------------------------------
\subsection{Measurement of the Rossiter-McLaughlin effect}
%---------------------------------------------------------------------------------------
As an important result, the sign of the RV anomaly 
shows that the orbital motion is set in the same direction as the stellar spin, similar to
the four previous observed RM effects on transiting exoplanets (\citet{Queloz00a,Winn05} for
HD209458; \cite{Winn06} for HD189733; \cite{Wolf07} for HD149026; and \cite{Narita07} for TrES-1).
The orbital inclination angle $i$ we derived is in full agreement with the value
determined by B07. The projected rotation velocity of the star $v\sin{I_s}$ 
determined from the RM fit is 22.9$\pm$1.2~{\kms}. 
This value is greater than our spectroscopic
determination from SOPHIE CCFs, and 
2-$\sigma$ greater than the determination of B07. However, \cite{Winn05} showed that 
$v\sin{I_s}$ measured with the analytical formulae from \cite{Ohta05} 
is biased toward larger velocities by approximately 10$\%$. Moreover, as the planet crosses the star at
the equatorial plane, assuming a differential rotational velocity of the star, the fitted $v\sin{I_s}$
corresponds to the maximum value. \cite{Kuker05} show that a differential rotational rate can be
as high as about 10$\%$ for a F8-type star.
The projected angle between the stellar spin axis and the normal of the orbital plane,
 $\lambda = 0.2\pm12.5\degr$, is consistent with complete alignment of the stellar spin 
and orbital angular momentum.

We extended our analysis in the search for the global minimum $\chi^2$, without any constraint on
the $v\sin{I_s}$. The result obtained gives a significantly larger $v\sin{I_s}$ value (29.5
$^{+3.1}_{-2.2}$ {\kms}) and a consistent lambda value (5.0$^{+17.8}_{-6.0}$ $\degr$). 
However, such a $v\sin{I_s}$ value is clearly incompatible with the
spectroscopic determination of the rotational velocity of the star.

\begin{figure}
\centering{
\includegraphics[scale=0.4]{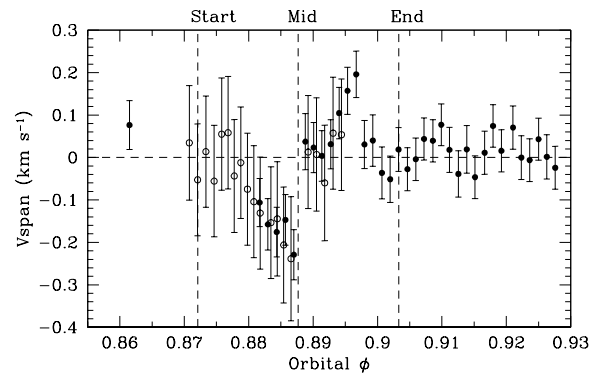}
\caption{\label {bisectortransit} Line-bisectors signature during
the spectroscopic transit as a function of the orbital phase. As in Fig.~\ref{rossiterfit} 
the filled and open circles represent the RV measurements obtained during the
first and second sequences of transit observations with SOPHIE, respectively.} }
\end{figure}

We also computed the line-bisector behavior following the procedure
described in \cite{Santos02}. As shown in
Fig.~\ref{bisectortransit}, a specific signature of the line-bisectors
can be found during the transit and is anti-correlated with the RVs
due to the fact that the crossing planet mainly affects the bottom of
the spectral lines.

Assuming the B07 stellar mass and period, of $1.298^{+0.062}_{-0.098} M_{\odot}$ and 5.63341 days, 
and using the fitted ratios $R_p/R_s$ and $a/R_s$, we also provide a new determination of the 
system parameters. Those include: orbital semi-major axis, $a$, stellar radius, $R_s$, and 
the planetary radius, $R_p$, mass, $M_p$, density $\rho_p$, and surface gravity $g_p$.
Our results are similar to those of B07 with smaller errors.
The RVs describing the spectroscopic transit are quite noisy and some of the measured RVs
(around orbital phase 0.885) present an unexpected RV shift. This shift does not seem to be due to 
an instrumental deviation and may be explained by a guidance decentring of the telescope.
However, the amplitude of the Rossiter anomaly is large enough that we can still
estimate the parameters of this system with quite good uncertainties.

\begin{table}[h]
\caption{\label {orbittable} Refined parameters of the Keplerian orbital solution
and the parameters of the HAT-P-2 system, compared to those determined in \cite[][B07]{Bakos07}. 
T$_0$ refers to the periastron time. The fitted parameters are
presented at the top of the table. The bottom part of the table lists the system parameters, derived
from the fitted ones. We present in the table both the spectroscopic determination and the result from
the fit of the star's $v\sin{I_s}$.}
\begin{center}{
\begin{tabular}{l l l}\hline
\hline
Parameter              & Value                               & Value from B07 \\\hline
Period [d] $^\star$        & 5.63341 (fixed)                     & 5.63341$\pm$0.00013\\
T$_0$ [HJD]            & 2454213.4794 $^{+0.0053}_{-0.0030}$ & 2454213.369 $\pm$ 0.041\\
e                      & 0.5163 $^{+0.0025}_{-0.0023}$       & 0.520 $\pm$ 0.010\\
$\omega$ [deg]         & 189.92 $^{+1.06}_{-1.20}$                    & 179.3 $\pm$ 3.6\\
K [\ms]                & 966.9 $\pm$ 8.3                     & 1011 $\pm$ 38\\
V$_0$ [\ms]            & -19855.1 $\pm$ 5.8                  & N.A.\\
$R_p/R_s$              & 0.06891 $^{+0.00090}_{-0.00086}$    & 0.0684 $\pm$ 0.0009\\   
$a/R_s$                & 10.28 $^{+0.12}_{-0.19}$            & $9.77^{+1.10}_{-0.02}$\\
$\lambda$ [deg]        & 0.2 $^{+12.2}_{-12.5}$              & N.A.\\
$v\sin{I_s}$ $_{RM}$ [\kms] & 22.9 $^{+1.1}_{-1.2}$          & N.A.\\ 
$v\sin{I_s}$ $_{Spectro}$ [\kms] & 21.3 $\pm$ 1.3            & 19.8 $\pm$ 1.6\\
i [deg]                & 90.0 $^{+0.85}_{-0.93}$                     & $> 84.6\ (2\sigma$)\\
$\Delta v_{KL}$ [\ms]  & -328 $\pm$ 41                       & -380 $\pm$ 35\\
$\Delta v_{SK}$ [\ms]  & -19584 $\pm$ 17                    & N.A.\\
$\Delta v_{S2}$ [\ms]  & -27 $\pm$ 12                     & N.A.\\
Stellar jitter [\ms] $^{\star\star}$  & 17                                  & 60\\\hline

[Fe/H] (dex) 	       & 0.11 $\pm$ 0.10		     & 0.12 $\pm$ 0.08\\
log $R'_{HK}$	       & -4.75 $\pm$ 0.02	             & -4.72 $\pm$ 0.05\\
$M_s$ [M$_{\odot}$]$^\star$ & 1.298 $^{+0.062}_{-0.098}$         & $1.298^{+0.062}_{-0.098}$\\
$a$ [AU]               & 0.0677 $^{+0.0011}_{-0.0017}$       & 0.0677 $\pm$ 0.0014\\
$R_s$ [\RS]            & 1.416 $^{+0.04}_{-0.062}$          & $1.474^{+0.042}_{-0.167}$\\
$M_p$ [\MJ]            & 8.62 $^{+0.39}_{-0.55}$             & 9.04 $\pm$ 0.50\\
$R_p$ [\RJ]            & 0.951 $^{+0.039}_{-0.053}$          & $0.982^{+0.038}_{-0.105}$\\
$\rho_p$ [g cm$^{-3}$] & 12.5 $^{+2.6}_{-3.6}$            & $11.9^{+4.8}_{-1.6}$\\
$g_p$ [m s$^{-2}]$     & 237 $^{+30}_{-41}$             & $227^{+44}_{-16}$\\
\hline
$^*$ Adopted from B07\\
$^{**}$ Short term jitter
\end{tabular}}
\end{center}
\end{table}

%=======================================================================================
\section{Discussion and Conclusion}
%=======================================================================================
The value of HAT-P-2b radius (0.95 \RJ) puts this object in the mass-radius 
diagram as an intermediate case
between Hot-Jupiters and low-mass stars. Its mean density of 12.5~g~cm$^{-3}$ 
is in between the Hot-Jupiter density (0.34 - 1.34 g cm$^{-3}$) and the density 
of the smallest transiting M dwarfs
OGLE-TR-122 and OGLE-TR-123 \citep{Pont05,Pont06}, which are 75 and 51 g cm$^{-3}$, respectively. A
similar paper \citep{Winn07} published results that are consistent with our
conclusions. Super-massive planets like
HAT-P-2b may constitute a new class of stellar companions, in between
Hot-Jupiters and low-mass stars and near the Brown dwarf population. HAT-P-2b is the first super-massive 
object around a F8 star for which the
exact mass has been determined. Such a massive close-in planet is not in agreement 
with the type II migration mechanism, which appears to be more efficient for planets around low-mass
stars \citep{Ida05}. This could suggest a different formation process for this object, such as 
fragmentation, or interactions between the planet and another companion or between 
the planet and the disk in the evolution process.

A system with an elliptic orbit is expected to move toward
pseudo-synchronization, with the stellar angular rotation velocity
tuned to near the angular velocity of the companion at periastron passage
\citep{Zahn77}. Given the Keplerian orbital
parameters, we computed the angular planetary speed at the periastron
position and found that a stellar radius of 1.42 {\RS} implies a
pseudo-synchronization rotational velocity $v\sin{I_s}$ of about 40~{\kms}
\citep[][equation 43]{Hut81}. Because the observed rotational velocity of the star is only about 21
{\kms}, the star is definitely not pseudo-synchronized to the planetary orbit. \cite{Peale99} 
indicated that the alignment and the synchronization 
timescales are of the same order of magnitude. Therefore, the 
lack of pseudo-synchronization indicates that the system was formed 
with the stellar spin aligned to the orbital angular momentum.

\acknowledgements{Part of these observations have been funded by the Optical Infrared Coordination network (OPTICON), a major international collaboration supported by the Research Infrastructures Programme of the European Commission's Sixth Framework Programme. 
N.C.S. would like to thank the support from Fundação para a
  Ciência e a Tecnologia, Portugal, in the form of a grant (reference
  POCI/CTE-AST/56453/2004). This work was supported in part by the EC's
  FP6 and by FCT (with POCI2010 and FEDER funds), within the HELAS
  international collaboration. A.E. would like to thank the support from the Swiss
  National Science Foundation (SNSF), Switzerland, in the form
  of a grant (reference PBSK2--114688). We thank the technical team from OHP who worked on the
  instrument SOPHIE and for their exceptional work.}

\bibliographystyle{aa}
\bibliography{biblio}

\end{document}